The relationship between individual variation in macroscale functional gradients and distinct aspects of ongoing thought.

Brontë Mckeown[a], Will H Strawson[b], Hao-Ting Wang[c], Theodoros Karapanagiotidis[a], Reinder Vos de Wael[d], Oualid Benkarim[d], Adam Turnbull[a], Daniel Margulies[e], Elizabeth Jefferies[a], Cade McCall[a], Boris Bernhardt[d] and Jonathan Smallwood[a].

[a] Department of Psychology/ York Neuroimaging Centre, University of York, United Kingdom
[b] Neuroscience, Brighton and Sussex Medical School, University of Sussex, United Kingdom
[c] Sackler Centre for Consciousness Studies, University of Sussex, United Kingdom
[d] McConnell Brain Imaging Centre, Montreal Neurological Institute and Hospital, McGill University, Montreal, QC, Canada
[e] Frontlab, Institut du Cerveau et de la Moelle épinière, UPMC UMRS 1127, Inserm U 1127, CNRS UMR 7225, Paris, France

Email and address for correspondence: bronte.mckeown@york.ac.uk; Department of Psychology, University of York, YO10 5DD

**Abstract**

Contemporary accounts of ongoing thought recognise it as a heterogeneous and multidimensional construct, varying in both form and content. An emerging body of evidence demonstrates that distinct types of experience are associated with unique neurocognitive profiles, that can be described at the whole-brain level as interactions between multiple large-scale networks. The current study sought to explore the possibility that whole-brain functional connectivity patterns at rest may be meaningfully related to patterns of ongoing thought that occurred over this period. Participants underwent resting-state functional magnetic resonance imaging (rs-fMRI) followed by a questionnaire retrospectively assessing the content and form of their ongoing thoughts during the scan. A non-linear dimension reduction algorithm was applied to the rs-fMRI data to identify components explaining the greatest variance in whole-brain connectivity patterns, and ongoing thought patterns during the resting-state were measured retrospectively at the end of the scan. Multivariate analyses revealed that individuals for whom the connectivity of the sensorimotor system was maximally distinct from the visual system were most likely to report thoughts related to finding solutions to problems or goals and least likely to report thoughts related to the past. These results add to an emerging literature that suggests that unique patterns of experience are associated with distinct distributed neurocognitive profiles and highlight that unimodal systems may play an important role in this process.

# 1 Introduction

When unoccupied by events in the immediate environment, such as during the so-called resting-state, humans often spend substantial amounts of time focused on information that is relevant to themselves but absent from the here and now. These self-generated experiences can be a source of unhappiness and distress (Killingsworth & Gilbert, 2010; Poerio et al., 2013). However, they can also allow individuals to mentally reframe their goals in a more concrete way (Medea et al., 2018), and reduce loneliness (Poerio et al., 2015), perhaps because of links between self-generated thought with creativity (Baird et al., 2012; Gable et al., 2019; Smeekens & Kane, 2016; Wang et al., 2018), social problem solving (Ruby et al., 2013), or generation of information based on semantic knowledge (Wang et al., 2019). Understanding the neural basis of these different patterns of ongoing thoughts, is therefore an important goal for cognitive neuroscience because it may help describe the underlying neural architecture which supports aspects of human cognition that are both beneficial and detrimental to health and well-being. In this study we examined whether an individual's ongoing thought patterns could predict individual variation in their functional organization at rest.

Contemporary views on how the structure of the cortex constrains its functions have identified the important roles that macroscale patterns of cortical organization play in determining cognition (Mesulam, 1998, Margulies et al., 2016). These patterns, or motifs, can be well captured by dimension reduction techniques that identify low-dimensional manifold spaces, often referred to as 'cortical gradients'. This approach has been important in characterising the axis upon which cortical structure is organised (Paquola et al., 2019; Vazquez-Rodriguez et al., 2019), how the specific topological features of the cortex give rise to different functional hierarchies (Margulies et al., 2016), describing changes in brain function in developmental disorders (Hong et al., 2019) and across primate species (Xu et al., 2019) and capturing dynamic changes between states of external task focus and self-generated social episodic thought (Turnbull et al., in press). One advantage of gradient approaches to neural function is that they describe multivariate whole-brain patterns of organization (i.e. the relationship between different neural systems) and so allow the investigation of whether macroscale features of cortical organization relate to features of cognition. This approach is particularly useful for understanding features of higher-order cognition which are hypothesised to depend upon the interaction between multiple neural systems (e.g. Smallwood et al., 2011; Smallwood & Schooler, 2015; Jefferies et al., 2020).

Our current study, therefore, explores the possibility that macroscale properties of the cortex captured by low-dimensional descriptors of functional organization at rest are related to individual variation in ongoing experience that emerge during this period. Resting-state fMRI was used to record patterns of intrinsic neural activity in a large cohort (N=277). We employed the BrainSpace toolbox (Vos de Wael et al., 2019) to calculate the dimensions that characterise the functional connectivity of the brain at rest. At the end of the scan, participants completed a questionnaire that retrospectively assessed their experiences during the scan. The questions were based on those used in previous studies exploring population variation in functional connectivity and aimed at capturing the heterogeneity of ongoing thought (Karapanagiotidis et al., 2017; Smallwood et al., 2016). While retrospective experience-sampling sacrifices temporal specificity, it is particularly beneficial for understanding the neural basis of ongoing experience because the absence of interruptions ensures that neural dynamics unfold in a relatively natural way (Smallwood & Schooler, 2015). Using these data, we examined whether specific types of thought measured at the end of the scan were predictive of individual variation along low-dimensional gradients of macroscale functional connectivity at rest. These data have previously been examined by Karapanagiotidis et al. (2019) who applied Hidden Markov modelling to identify neural states occurring at rest. They found states linked to autobiographical planning and intrusive rumination that were related to differences in the relative dominance of frontoparietal and motor systems, and default mode and visual systems.

Prior studies have highlighted three cortical gradients which each relate to meaningful features of cognition. The first gradient describes the difference between regions of unimodal and transmodal cortex (Margulies et al., 2016). Studies have shown that this neural motif is observed when participants must use information from memory to guide behaviour, such as when visuospatial decisions must be made with previously encountered information rather than immediate perceptual information (Murphy et al., 2018, 2019). The second gradient is related to the dissociation between unimodal systems concerned with vision and sensorimotor systems (Margulies et al., 2016). Finally, the third gradient describes a distinction between the so-called default mode and task-positive systems. This pattern is often observed when researchers compare easy and demanding cognitive tasks (Cole et al., 2013; Duncan, 2010). Prior studies have shown that this pattern is linked to the difference between on and off task states and that this distinction also helps describe neurocognitive changes related to the passage of time (Turnbull et al., in press). Our study aimed to explore whether any of these macroscale neural motifs were related to the participants reports at the end of the experimental session.

## 2 Methods

### 2.1 Participants

Two hundred and seventy-seven healthy participants were recruited from the University of York. Written informed consent was obtained for all participants and the study was approved by the York Neuroimaging Centre Ethics Committee. Twenty-three participants were excluded from analyses; two due to technical issues during the neuroimaging data acquisition and twenty-one for excessive movement during the fMRI scan (mean framewise displacement (Power et al., 2014) > 0.3 mm and/or more than 15% of their data affected by motion), resulting in a final cohort of n = 254 (169 females, mean ± SD age = 20.7±2.4 years). The questionnaire and functional MRI data in this study are the same as those reported in Karapanagiotidis et al. (2019).

### 2.2 Data and Code availability

Gradient maps one to ten from the group-averaged dimension reduction analysis described in section 2.5.3 below are publicly available on NeuroVault in a collection with the title of this article (https://neurovault.org/collections/6746/). Raw fMRI and questionnaire data are restricted in accordance with ERC and EU regulations. All code used in the production of this manuscript is publicly available online in the following repository: https://github.com/Bronte-Mckeown/GradientAnalysis.

### 2.3 Retrospective experience-sampling

Participants' experience during the resting-state fMRI scan was sampled by asking them to retrospectively report their thoughts during the resting-state period at the end of the scan. Experience was measured using a 4-point Likert scale with the question order randomised (all 25 questions are shown in Table 1).

**Table 1.** 25-item experience-sampling questionnaire completed at the end of the resting-state fMRI scan. Answers were given on a 4-point Likert scale ranging from "Not at all" to "Completely".

| Dimension | Question (My thoughts…) |
|---|---|
| Vivid | … were vivid as if I was there |
| Normal | … were similar to thoughts I often have |
| Future | … involved future events |
| Negative | … were about something negative |
| Detail | … were detailed and specific |
| Words | … were in the form of words |
| Evolving | … tended to evolve in a series of steps |
| Spontaneous | … were spontaneous |
| Positive | … were about something positive |
| Images | … were in the form of images |
| People | … involved other people |
| Past | … involved past events |
| Deliberate | … were deliberate |
| Self | … involved myself |
| Stop | … were hard for me to stop |
| Distant time | … were related to a more distant time |
| Abstract | … were about ideas rather than events or objects |
| Decoupled | … dragged my attention away from the external world |
| Important | … were on topics that I care about |
| Intrusive | … were intrusive |
| Problem Solving | … were about solutions to problems (or goals) |
| Here and Now | … were related to the here and now |
| Creative | … gave me a new insight into something I have thought about before |
| Realistic | … were about an event that has happened or could take place |
| Same Theme | … at different points in time were all on the same theme |

## 2.4 Procedure

All participants underwent a 9-minute resting-state fMRI scan. During the scan, they were instructed to passively view a fixation cross and not to think of anything in particular.

Immediately following the scan, they completed the 25-item experience-sampling questionnaire while still in the scanner.

## 2.5 Resting-state fMRI

### 2.5.1 MRI data acquisition

MRI data were acquired on a GE 3 Tesla Signa Excite HDxMRI scanner, equipped with an eight-channel phased array head coil at York Neuroimaging Centre, University of York. For each participant, we acquired a sagittal isotropic 3D fast spoiled gradient-recalled echo T1-weighted structural scan (TR = 7.8 ms, TE = minimum full, flip angle = 20°, matrix = 256x256, voxel size = 1.13x1.13x1 mm3, FOV = 289x289 mm2). Resting-state fMRI data based on blood oxygen level-dependent contrast images with fat saturation were acquired using a gradient single-shot echo-planar imaging sequence (TE = minimum full (≈19 ms), flip angle = 90°, matrix = 64x64, FOV = 192x192 mm2, voxel size = 3x3x3 mm3, TR = 3000 ms, 60 axial slices with no gap and slice thickness of 3 mm). Scan duration was 9 minutes which allowed us to collect 180 whole-brain volumes. These acquisition details are identical to the ones described in Karapanagiotidis et al. (2019).

### 2.5.2 MRI data pre-processing

fMRI data pre-processing was performed using SPM12 (http://www.fil.ion.ucl.ac.uk/spm) and the CONN toolbox (v.18b) (https://www.nitrc.org/projects/conn) (Whitfield-Gabrieli & Nieto-Castanon, 2012) implemented in Matlab (R2018a) (https://uk.mathworks.com/products/matlab). Pre-processing steps followed CONN's default pipeline and included motion estimation and correction by volume realignment using a six-parameter rigid body transformation, slice-time correction, and simultaneous grey matter (GM), white matter (WM) and cerebrospinal fluid (CSF) segmentation and normalisation to MNI152 stereotactic space (2 mm isotropic) of both functional and structural data. Following pre-processing, the following potential confounders were statistically controlled for: 6 motion parameters calculated at the previous step and their 1st and 2nd order derivatives, volumes with excessive movement (motion greater than 0.5 mm and global signal changes larger than z = 3), linear drifts, and five principal components of the signal from WM and CSF (CompCor approach) (Behzadi et al., 2007). Finally, data were band-pass filtered between 0.01 and 0.1 Hz. No global signal regression was performed. The pre-processing steps reported here are identical to the ones described in Karapanagiotidis et al. (2019).

### 2.5.3 Whole-brain Functional Connectivity: Dimension reduction

Following pre-processing, the functional time-series from 400 ROIs based on the 400 Schaefer parcellation (Schaefer et al., 2018) were extracted for each individual. A connectivity matrix for each individual was then calculated using Pearson correlation resulting in a 400x400 connectivity matrix for each participant. These individual connectivity matrices were then averaged to calculate a group-averaged connectivity matrix. The Brainspace Toolbox (Vos de Wael et al., 2019) was then used to extract ten group-level gradients from the group-averaged connectivity matrix (dimension reduction technique = diffusion embedding, kernel = normalized angle, sparsity = 0.9). Although we were only interested in the first three gradients as they all have reasonably well described functional associations, we extracted ten gradients to maximise the degree of fit between the group-averaged gradients and the individual-level gradients (see Inline Supplementary Table 1 for the average degree of fit for gradients one to three when extracting ten gradients compared to three). These group-averaged gradients act as a template to which individual gradients can be compared, to allow an investigation of individual differences along each gradient in the current sample. The variance explained by each group-averaged gradient one to ten is shown in Inline Supplementary Figure 1.

The group-level gradient solutions were aligned using Procrustes rotation to a subsample of the HCP dataset ([n=217, 122 women, mean $\pm$ sd age = 28.5 $\pm$ 3.7 y]; for full details of subject selection see Vos de Wael et al. (2018)) openly available within the Brainspace toolbox (Vos de Wael et al., 2019). This alignment step improves the stability of the group-level gradient templates by maximising the comparability of the solutions to those from the existing literature (i.e. Margulies et al., 2016). The first three group-averaged gradients, with and without alignment to the HCP data are shown in Inline Supplementary Figure 2. To demonstrate the benefits of this alignment step, we calculated the similarity using Spearman Rank correlation between the first five aligned and unaligned group-level gradients with the first five gradients reported in Margulies et al. (2016) which were calculated using 820 participants over an hour resting-state scan. Aligning our gradients with a subsample of the HCP data increased the similarity between our gradients and Margulies' et al (2016) gradients (see Inline Supplementary Table 2).

Using identical parameters, individual-level gradients were then calculated for each individual using their 400x400 connectivity matrix. These individual-level gradient maps were aligned to the group-level gradient maps using Procrustes rotation to improve comparison between the group-level gradients and individual-level gradients (N iterations = 10). This analysis resulted in ten group-level gradients and ten individual-level gradients for each participant explaining maximal whole-brain connectivity variance in descending order. All ten group-level gradients are shown in Figure 1, however, only the first three gradients were retained for further analysis. To demonstrate the variability of individual-level gradients, Inline Supplementary Figure 3 shows the highest, lowest and median similarity gradient maps for gradients one to three.

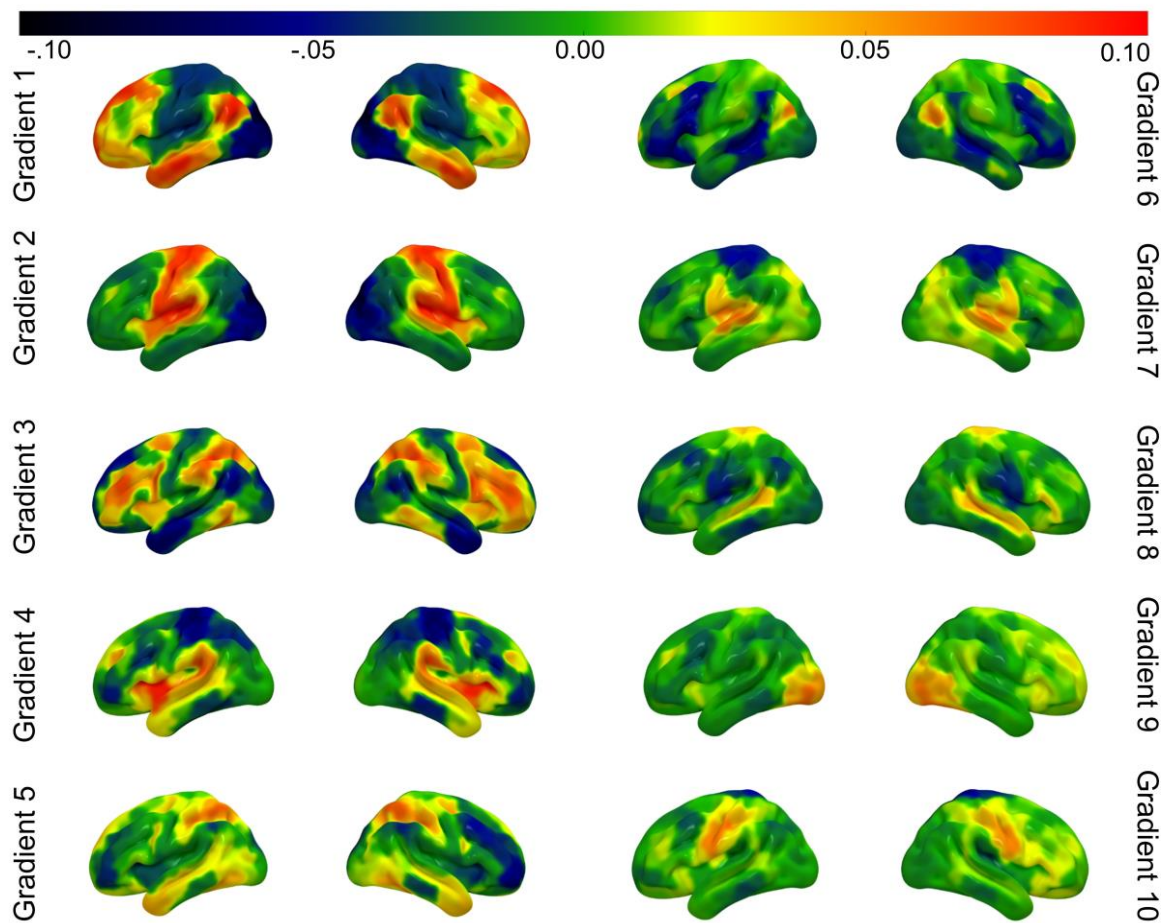

**Figure 1.** *Group-averaged gradients one to ten (left and right lateral views) explaining maximal variance in whole-brain connectivity patterns.* Regions that share similar connectivity profiles fall together along each gradient (similar colours) and regions that have more distinct connectivity profiles fall further apart (different colours). The positive and negative loading is arbitrary. Regions which fall at the extreme end of each gradient have the greatest dissimilarity in their connectivity profiles. Only gradients one to three were included in the multivariate analysis. These ten group-averaged gradient maps are publicly available on NeuroVault (https:/neurovault.org/collections/6746/).

### 2.5.4 Individual-level Similarity Analysis: Spearman's Rank Correlation

In order to investigate individual differences for each of the three connectivity gradients, a Spearman's rank correlation was used to calculate the extent to which each individual-level gradient was related to each group-level gradient. In this way, the correlation coefficient calculated for each participant for gradients one to three is used as a second-order statistic indicating the similarity between the group-level and individual-level gradients. Fishers R-to-Z transformation was applied to these correlation coefficient scores. These z-transformed regression coefficients will be referred to as 'gradient similarity scores' from this point onwards. These similarity scores were then entered as dependent variables in subsequent multivariate regression analyses to investigate whether individual variation in ongoing thought patterns could predict individual variation along the first three whole-brain connectivity gradients. A schematic for the analysis pipeline is shown in Figure 2.

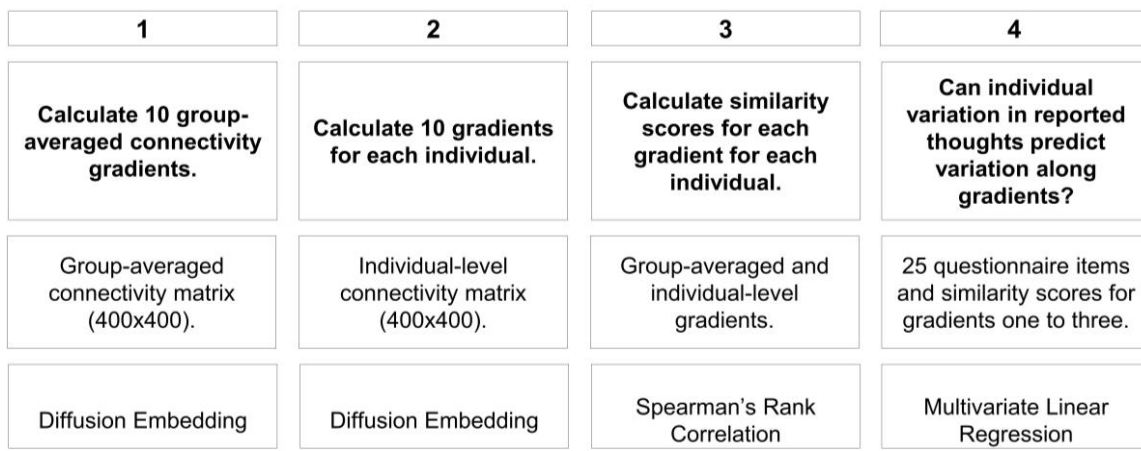

**Figure 2.** *Summary of the analysis pipeline.* Numbers represent order of step. Top panel in bold describes the overarching goal of each step. Middle panel specifies the data being used. Bottom panel indicates which analysis or statistical test was used to achieve the step.

## 3 Results

### 3.1 Experience-sampling responses

The experience-sampling data is summarised in figure 3, revealing the distribution of responses for each item as well as the covariance between each item. While some questionnaire items are significantly correlated, the variance inflation factor for each questionnaire item was <2, indicating that multicollinearity is not a concern in the multivariate regression analysis described below.

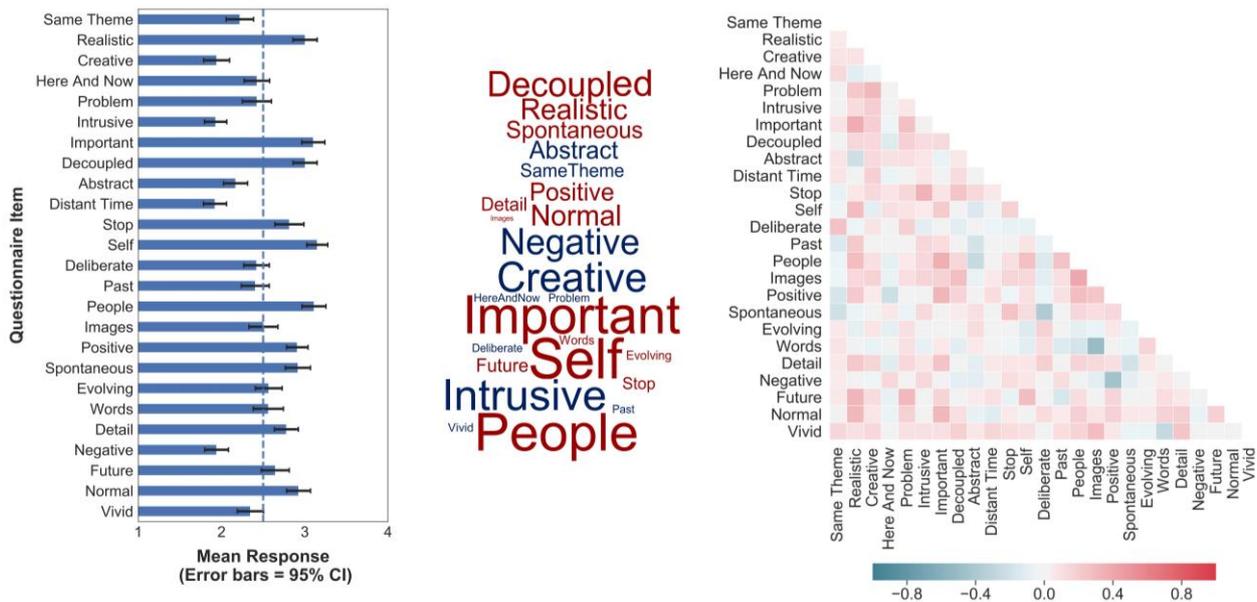

**Figure 3**. *Summary information describing the distribution of the retrospective measures of ongoing experience recorded in our study*. In the left-hand panel, the bar graph shows the average loading on each question relative to the mid-point of the scale (indicated by the dashed line). The error bars reflect 95% confidence intervals, adjusted to account for family-wise error (i.e. the 25 items). The word cloud shows this information in a different form in which the size of the word describes its distance from the mid-point and its colour (cold / warm) reflects its loading. The right-hand panel illustrates the patterns of covariation between these items (Pairwise Pearson correlation).

## 3.2 Multivariate analysis

We examined whether there was any relationship between the low-dimensional representations of the macroscale organization of neural function and the experience of participants during the scanning. We used a Multivariate linear regression (SPSS; version 26) in which individual items from the experience-sampling questionnaire were included as explanatory variables and the similarity scores for gradients one to three were entered as dependent variables. Age, gender and mean movement during the scan were entered as nuisance covariates. This analysis revealed that there was a multivariate effect for the 'problem-solving' item [Pillai's trace = .046, F (3, 223) = 3.54, p = .015] and the 'past' item [Pillai's trace = .051, F (3, 223) = 3.97, p = .009]. These results establish that these two aspects of the questionnaire varied significantly with the similarity scores for the functional motifs apparent at rest.

We calculated the parameter estimates for these multivariate effects linked to thoughts of the 'past' (Gradient one (b = -0.018, 95% CI = [-0.042, 0.006], p = .137), Gradient two (b = -0.032, 95% CI = [-0.056, -0.008], p = .009) and Gradient three (b = 0.006, 95% CI = [-0.011, 0.024], p = .490) and for 'problem-solving' (Gradient one (b = 0.020, 95% CI = [-0.005, 0.044], p = .112), Gradient two (b = 0.036, 95% CI = [0.011, 0.061], p = .004) and Gradient three (b = -001, 95% CI = [-0.019, 0.018], p = .951)). In both cases, therefore, the only association in which the error bars did not overlap with zero was with Gradient two.

Together these analyses revealed that the multivariate effect for the 'problem-solving' item is most clearly positively associated with gradient two while the multivariate effect for the 'past' item shows the reverse pattern. To understand these associations, we visualised the average map of gradient two for individuals in the top and bottom third of similarity with the group-level description, and also calculated the difference. This data is presented in the left-hand panel of Figure 4 where it can be seen that individuals with higher similarity to group-averaged gradient two showed decreased shared connectivity between the visual and sensorimotor systems.

To visualise the associations between the 'problem-solving' and 'past' questionnaire items with gradient two, we calculated the unique variance associated with gradient two and both questionnaire items separately. To do this, we calculated the residual variance linked to both types of thoughts using linear regressions in which the dependent variable was gradient two similarity scores and the explanatory variables were all of the questionnaire items (as well as age, gender and mean movement) except for the relevant item (either 'problem-solving' or 'past'). We performed a comparable analysis to identify the residual variance in gradient two. Together this data is presented in the right-hand panel of Figure 4 where it can be seen that individuals with high similarity scores for gradient two reported more problem-solving thoughts and fewer past-related thoughts.

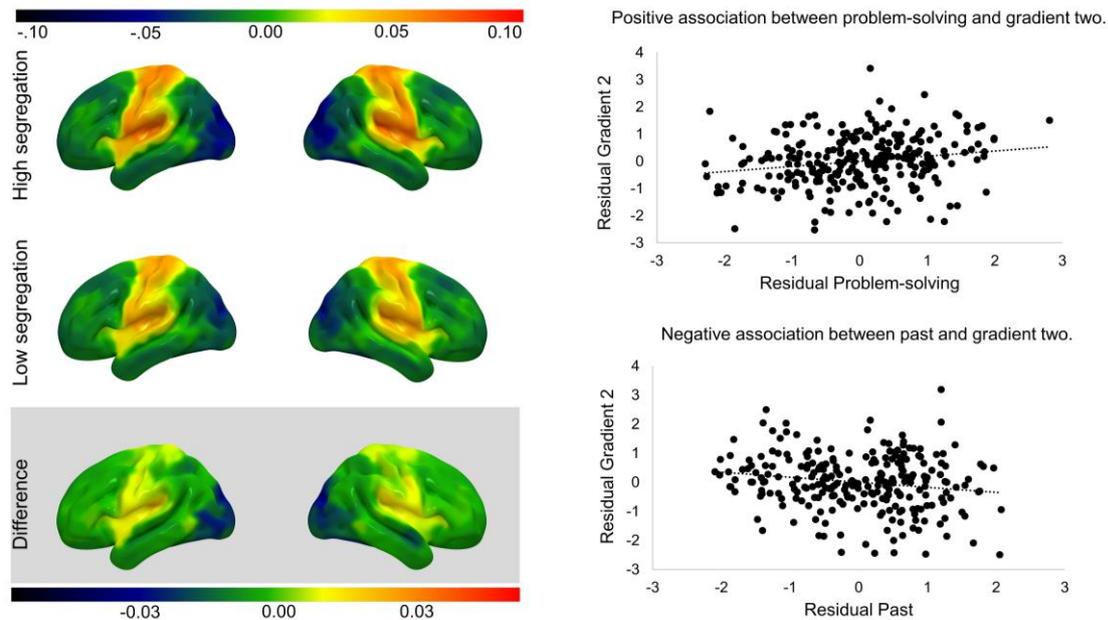

**Figure 4.** *Greater functional segregation between visual and sensorimotor cortices was positively associated with reports of problem-solving thoughts during rest and negatively associated with reports of thoughts about past events*. Left panel: group-averaged maps for high (top) and low (middle) similarity scores for gradient two as well as the difference between these groups (bottom). The top colour bar reflects the scale of the high and low similarity group-averaged maps while the bottom colour bar reflects the scale of the difference map. Individuals with high similarity scores showed more functional segregation between visual (blue) and sensorimotor cortices (orange). The proximity of colours reflects greater similarity in connectivity patterns between regions. Right panel (upper): Scatterplot of residuals describing the positive relationship between gradient two similarity and the 'problem-solving' questionnaire item. Each point is a participant. Right panel (lower): Scatterplot of residuals describing the negative relationship between gradient two similarity and the 'past' questionnaire item. Using raw scores, a Pearson correlation confirmed this negative the positive association with problem solving thoughts ($r(252) = .16$, $p = .013$) and a negative relationship with past related thoughts ($r(252) = -.13$, $p = .040$).

## 4 Discussion

The current study employed a data-driven approach to identify whole-brain connectivity patterns associated with distinct patterns of ongoing thought at rest. Specifically, we were interested in identifying whether three reasonably well-described macroscale patterns of neural function ('cortical gradients') were related to the experiences an individual had at rest. Participants completed a rs-fMRI scan followed by an experience-sampling questionnaire retrospectively assessing the content and form of their ongoing thoughts during the scan. To reduce the dimensional structure of the rs-fMRI data we used a non-linear dimension reduction algorithm to embed the functional connectivity in a low-dimensional space. We found that individuals with less similarity between the pattern of functional connectivity in visual and sensorimotor cortices were more likely to report thoughts related to finding solutions to problems or goals and less likely to report thoughts related to past events (as demonstrated in figure 4).

It is worth considering the relationship between the current results and previous findings reported by Karapanagiotidis et al. (2019). They used the same dataset as the current study and applied Hidden Markov modelling to identify neural states. This analysis found two states which were associated with measures of experience. One state was linked to patterns of autobiographical planning (future-oriented problem-solving) and was associated with the dominance of the motor system relative to the visual system. In contrast, a second state was linked to intrusive rumination about the past and exhibited reasonably similar levels of activity in both the visual and motor systems. There is therefore an encouraging correspondence between the results of the current analysis, which entails a decomposition of the resting-state data into low dimensional manifolds, and the prior analyses which identifies neural states which reoccur at rest.

Together, these results add to a growing body of evidence that suggest neural processing in either primary motor or visual cortex may play an important role in aspects of higher-order cognition, especially those that involve imagining events other than those in the immediate environment. For example, Medea and colleagues asked participants to complete two writing sessions in which they either wrote about three personal goals or three TV programmes (Medea et al., 2018). Before and after each writing session participants completed an experience-sampling session. They found that if participants reported future-directed thought between writing session one and two, the concreteness of their personal goals increased between sessions. Importantly, this pattern was most pronounced for individuals who showed stronger connectivity between the hippocampus and a region of motor cortex at rest. Consistent with the possibility that motor cortex activity can be important during periods of self-generated thought, Sormaz and colleagues used online experience-sampling and found that neural patterns in regions of motor cortex were able to differentiate between thoughts related to a working memory task and those related to personal concerns about the future (Sormaz et al., 2018). Matheson and Kenett (2020) propose that the motor system is likely to be important in creative problem solving because of the capacity for this system to model the simulation of possible actions. Future work will be needed to understand the precise role that motor cortex activity plays in different patterns of ongoing thought.

There is also converging evidence from fMRI studies suggests that primary visual cortex is recruited during internal processing independent from external stimuli (Muckli, 2010). For example, activity in visual cortex has been observed during the retention period of a working memory task in which no external stimulus was presented (Harrison & Tong, 2009), while Japardi et al., (2018) found that visual system connectivity was important during periods of creativity for visual artists. Furthermore, Villena-Gonzalez et al. (2018) found that the degree of connectivity between the visual cortex and regions of posterior medial cortex were associated with a tendency to employ social information when engaged in task-based prospection. Together with these prior studies, the current work provides converging evidence linking processes in unimodal cortex to aspects of imaginative thought, an important question for future work to explore.

More generally our data suggests that different aspects of ongoing thought may vary in the degree to which unimodal systems are integrated. Mesulam (1998) argued that if a cortical system only contained unimodal regions, it would have difficulties in performing cognitive acts that depended on regularities that spanned multiple modalities. The connectivity pattern identified in gradient two recapitulates this theoretical functional organization proposed by Mesulam; the relative segregation of the unimodal systems coupled with common connectivity with transmodal and integrative systems such as the default mode network (See figure 5 for a schematic of this architecture). It is possible that the degree of integration between these unimodal systems may help encode and retrieve visual and auditory features of an experience, a process for which regions in the medial temporal lobe such as the hippocampus (Moscovitch et al., 2016) or the anterior temporal lobe (Ralph et al., 2017) may be particularly important. Based on our data we hypothesise that different types of experience may vary with the degree of overlap between these primary systems. Plausibly, a focus on thoughts relating to the past can rely on co-recruitment in both visual and motor regions because these experiences can capitalise on pre-existing memory traces and which may have been particularly strongly encoded if they spontaneously come to mind in a fluent fashion. In contrast, when attempting to generate a novel solution to a problem, it is less easy to capitalise directly on whole-brain associations from the past. Problem solving, therefore, may depend to a greater extent on processes that simulate the specific sequence of actions that should be performed, or, the arrangement of specific features of the environment, and which may be relatively achievable without interactions across different forms of unimodal cortex.

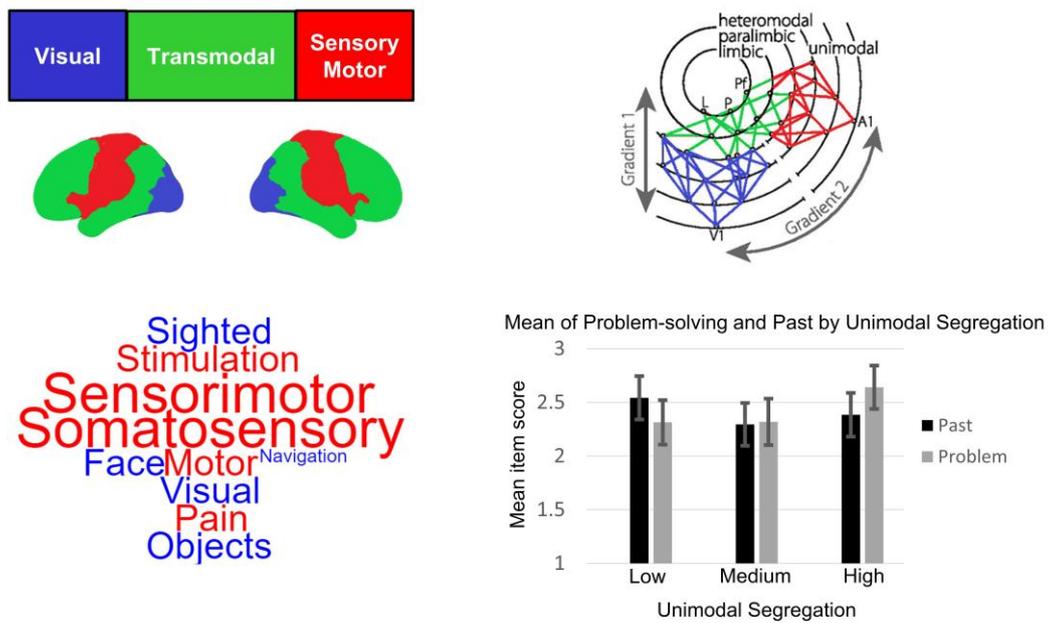

**Figure 5.** *Schematic of a hypothesised relationship between the macroscale organization and patterns of thought with different features.* Left panel (top): Simplified schematic of gradient two representing the segregation of unimodal systems with intermediary transmodal regions in between. Left panel (bottom): Word clouds representing the Neurosynth terms associated with the positive (red) and negative (blue) end of gradient two demonstrating the differences in function in the different unimodal systems. Font size represents the magnitude of the relationship, while the colour illustrates the associated system (blue = visual and red = sensorimotor). Right panel (top): Modified illustration of Mesulam's (1998) proposal of how the cortex is organised according to a functional hierarchy of processing from distinct unimodal systems to integrative transmodal regions. Gradient 1 and 2 labels correspond to the results reported in Margulies et al. (2016). Right panel (bottom): Schematic illustration of how unimodal segregation and integration may be differentially associated with distinct aspects of experience. We divided individuals into low, medium and high groups based on the similarity between visual and sensorimotor systems and plotted the mean scores for problem-solving and past related thoughts. It can be seen that based on our data individuals showing less segregation between unimodal systems reported more thoughts about past events and fewer problem-solving thoughts (and vice versa). Error bars indicate the 95% confidence intervals.

Finally, the current results lend further support to the view that it is necessary for researchers to distinguish between distinct types of ongoing thought (Seli et al., 2018). Our study shows that different types of ongoing thought are differentially associated with macroscale connectivity patterns, suggesting that different types of ongoing thought are supported by related but distinct mechanisms. Previously, many researchers have conflated various types of ongoing thought under one unitary measure (e.g. Mason et al., 2007; Smallwood et al., 2008). The current results suggest that in doing so, researchers may have made erroneous conclusions regarding the neural correlates of states that may often be discussed together under broad umbrella concepts such as 'mind-wandering'. Accordingly, our results demonstrate the value of the family-resemblances view of mental states which stresses the importance of operationalizing and describing the specific type of experience under investigation (Seli et al., 2018).

Although our study highlights a relationship between the macroscale organization of neural function at rest and concurrent patterns of ongoing experience, it nonetheless leaves several important questions unanswered. First, the present study focused on assessing static rather than dynamic functional connectivity and so cannot address important features of the relationship between neural dynamics and experience (Kucyi, 2018; Lurie et al., 2018). The choice of static functional connectivity coupled with retrospective sampling at the end of the scan means that the current study is unable to identify neuro-experiential associations that are highly transient and dynamic. One way to extend the current findings could be to incorporate sliding window analysis which consists of calculating a given functional connectivity measure (e.g. correlation) over consecutive windowed sections of data and to measure experience on multiple occasions. This method results in a time series of functional connectivity values which can then be used to assess the temporal fluctuations in functional connectivity within a scanning session (Hutchison et al., 2013). Future work combining gradient analyses with dynamic functional connectivity techniques such as Hidden Markov modelling (Vidaurre et al., 2018) or time-varying multi-network approaches (Betzel & Bassett, 2017) with multiple online experience-sampling measures, could help understand how macroscale connectivity patterns and ongoing thought patterns fluctuate together over time.

While retrospective sampling was chosen in the current study to allow neural dynamics to unfold in a relatively natural way over the scan period (Smallwood & Schooler, 2015), this method is not without its limitations which are important to consider when interpreting the current results. For example, retrospective sampling, compared to online sampling, relies more heavily on the participant's ability to remember their own thoughts. This introduces a number of potential confounds such as participants only reporting their most salient thoughts over the scanning period or some participants being more able than others to accurately recall their own thoughts. However, it is important to note that with more frequent sampling of ongoing experience the time series upon which cortical gradients are calculated would be shortened and this could temper the reliability of these metrics as indicators of neural function (Hong et al., 2020). Another limitation of the current study is that there was no experimental manipulation, making the causal link between macroscale patterns of neural activity and ongoing thoughts unclear. This issue could be fruitfully explored by priming participants to think about finding solutions to problems or goals and observe the changes in ongoing neural connectivity, or through the use of techniques such as trans-magnetic stimulation to disrupt either visual or motor cortex and observe the subsequent changes in patterns of ongoing thought.

Finally, it is important to note that it is not necessarily the case that the absence of associations with the majority of the items in this battery indicates that these aspects of experience are unimportant at rest. It is possible that other types of neural metric that focus on local patterns are important (such as fractional amplitude of low-frequency fluctuations [fALFF] or regional homogeneity [ReHo]; for example, see Gorgolewski et al., 2014) and that these types of relationship would be missed by our current analytic approach which focused on macroscale patterns of neural organization. It is also possible that other features of analysis are more state-like and detecting these types of patterns would require the capacity to measure both ongoing experience and neural experience across several time points (see Vatansever et al., 2020 for an exploration of this question). Finally, although resting-state is a common method for acquiring brain data and one in which patterns of ongoing experience are important, it is also possible that other contexts provoke different types of experience (for example see Ho et al., 2020). Thus, while our study shows that patterns of problem solving and past related experience are likely to be important aspects of a participants experiences at rest, in the future it will be important to carefully determine the most appropriate items for efficiently describing different features of experience in different situations and examining their relationships to a range of different metrics of static and dynamic neural function.

## 5  Conclusions

The current study investigated whether individual variation in ongoing thought patterns is associated with low-dimensional representations of macroscale functional connectivity at rest. Results revealed that reports of thoughts about finding solutions to problems was linked to greater segregation between the visual and sensorimotor systems, while thoughts about past events was linked to less segregation. These associations suggest that the degree of segregation of unimodal systems determine important features of ongoing experience. Future work could investigate the extent to which priming individuals to think about particular topics changes patterns of ongoing neural activity, or, use neurostimulation techniques to alter neural activity and examine how this changes ongoing experience. Such studies would provide important causal evidence on the relationship between macroscale patterns of neural activity and patterns of ongoing thought. Moving forward, it is likely to be increasingly important for scientists studying patterns of functional connectivity in states such as rest, or even tasks to acquire measures of ongoing experience in order to fully appreciate the significance of neural motifs that are revealed through the application of advanced analysis methods. Likewise, it will be important for researchers studying patterns of ongoing thought to recognise that these states are sometimes encoded in complex distributed whole-brain pattern of neural activity, and are not always localizable to a specific modular region of cortex.

# 6 Funding

This project was supported by European Research Council Consolidator awarded to JS (WANDERINGMINDS–646927).

# 7 Role of funding source

The funding source was not involved in the study design, data collection, analysis or interpretation of data; in the writing of the report; or in the decision to submit the article for publication.

# 8 Declarations of interest

None

## 9 Supplementary materials

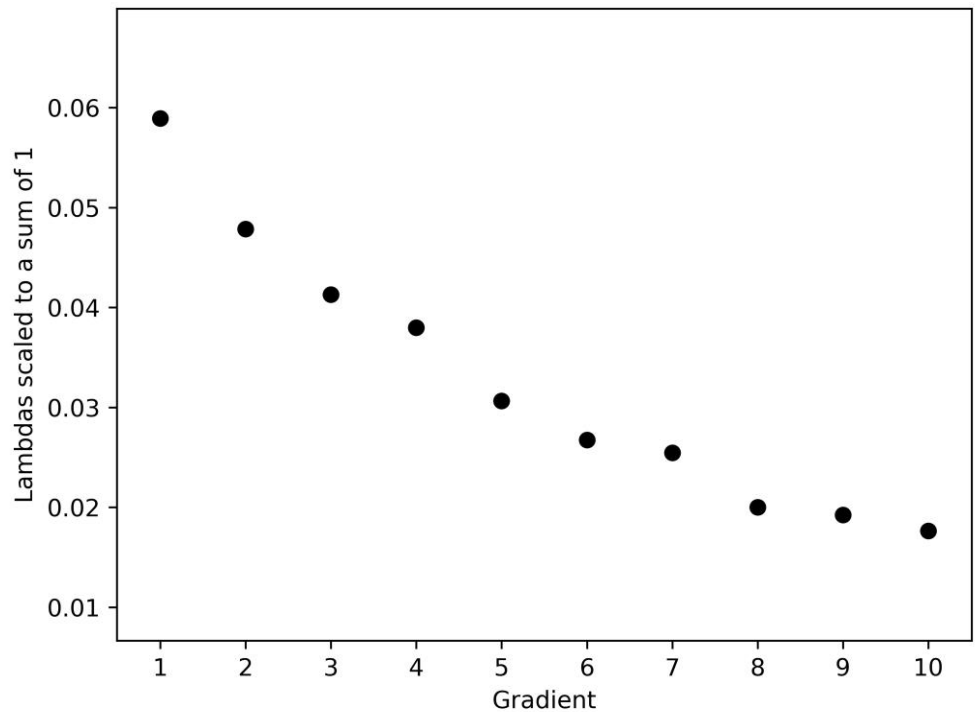

Scree plot of the scaled eigenvalues of the group-averaged gradients.

**Inline Figure S1.** *Scree plot showing the proportion of variance explained by each of the group-averaged whole-brain connectivity gradients one to ten.* Y-axis shows the eigenvalues scaled to a sum of 1. X-axis shows the gradient number. The first three gradients were retained for further multivariate analyses as these gradients have the clearest mapping to cognitive function (e.g. Murphy et al., 2018, 2019; Turnbull et al., in press).

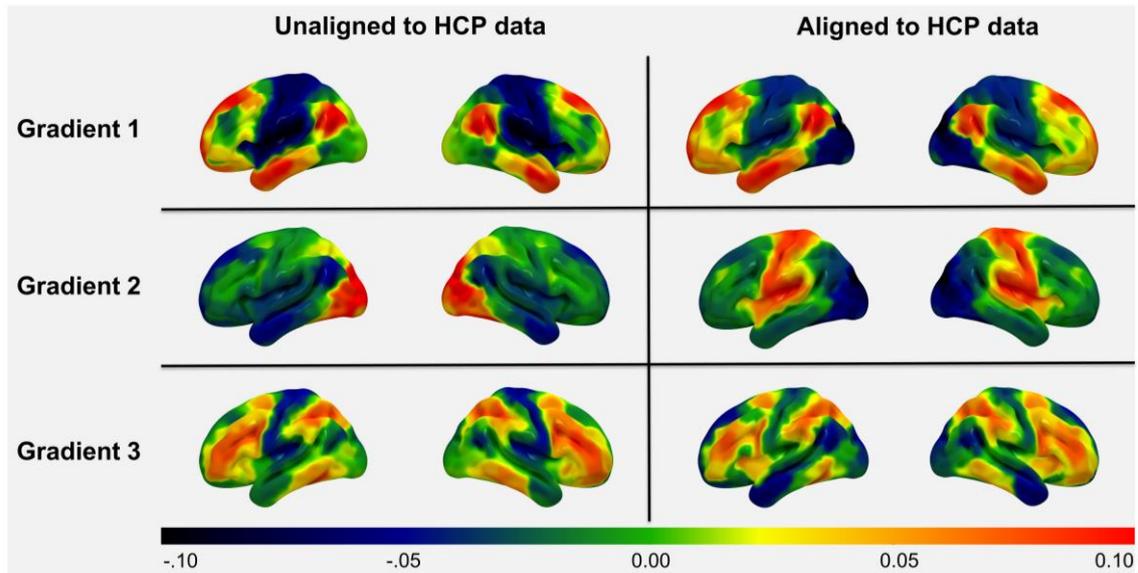

**Inline Figure S2.** *Demonstration of how aligning the group-level gradients to a subsample of the HCP dataset using Procrustes rotation changes the first three group-level gradients.* Regions that share similar connectivity profiles fall together along each gradient (similar colours) and regions that have more distinct connectivity profiles fall further apart (different colours). It is important to note that the positive and negative loading is arbitrary and can flip each time the diffusion embedding is applied to the data. For example, in this figure, the visual cortex along gradient two has a positive loading in the unaligned map but has a negative loading in the aligned map. Thus, differences in loadings are not meaningful and occur randomly.

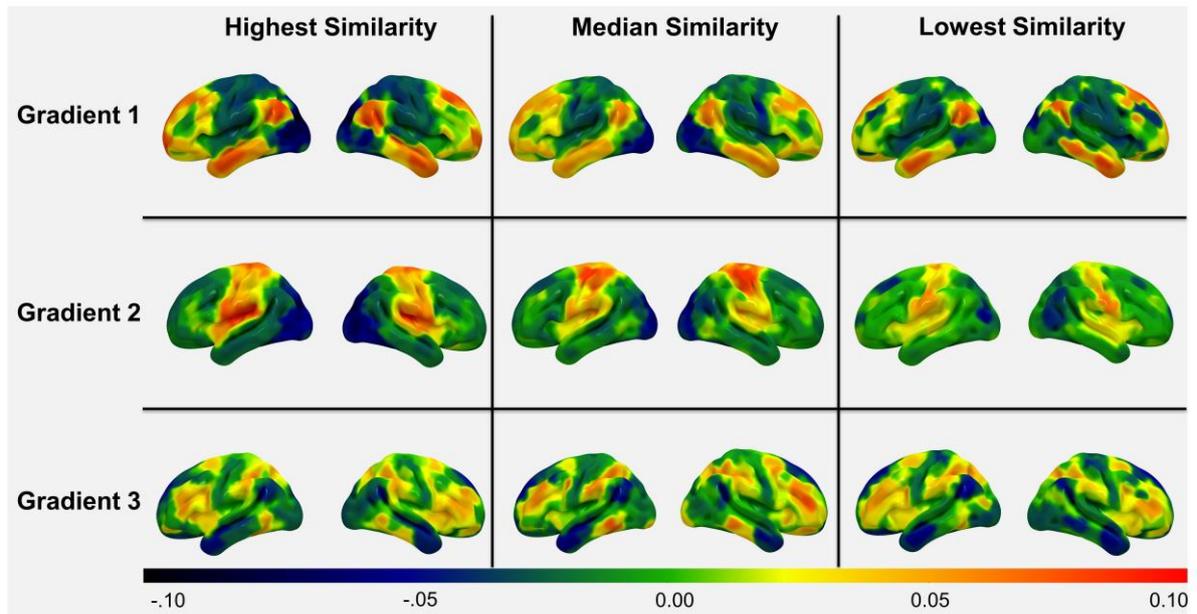

**Inline Figure S3.** *Individual-level connectivity gradients one to three which have the highest (left), median (middle) and lowest (right) similarity with the respective group-level gradients to demonstrate the variability of gradients across participants in the current sample.* Regions that share similar connectivity profiles fall together along each gradient (similar colours) and regions that have more distinct connectivity profiles fall further apart (different colours). The positive and negative loading is arbitrary.

**Inline Supplementary Table 1.** This table shows the improvement in the degree of fit (or similarity) between individual-level and group-level gradients when extracting ten gradients compared to only extracting three gradients. Mean similarity was calculated by averaging all participant's R-to-Z transformed Spearman Rank correlation coefficients for each respective gradient.

| Extracting 3 gradients: | Minimum | Maximum | Mean | Std. Deviation |
|---|---|---|---|---|
| Gradient 1 | 0.31 | 1.31 | 0.84 | 0.21 |
| Gradient 2 | 0.28 | 1.48 | 0.84 | 0.25 |
| Gradient 3 | -0.07 | 1.04 | 0.57 | 0.19 |
| **Extracting 10 gradients:** | **Minimum** | **Maximum** | **Mean** | **Std. Deviation** |
| Gradient 1 | 0.7 | 1.76 | 1.36 | 0.16 |
| Gradient 2 | 0.9 | 1.85 | 1.37 | 0.16 |
| Gradient 3 | 0.58 | 1.38 | 1.12 | 0.12 |

**Inline Supplementary Table 2.** Spearman rank correlation values for the first five aligned and unaligned group-level gradients with the first five group-level gradients reported in Margulies et al (2016). This demonstrates that aligning the group-level gradients to the subsample of HCP data improves correspondence between the gradients calculated in the current study and previous literature.

|  | **Aligned to HCP** | **Unaligned to HCP** |
|---|---|---|
| Gradient 1 | 0.62 | 0.4 |
| Gradient 2 | -0.47 | 0.23 |
| Gradient 3 | -0.45 | -0.38 |
| Gradient 4 | -0.2 | 0.07 |
| Gradient 5 | -0.18 | -0.03 |